\documentclass[a4paper,12pt,eqno]{article}
\usepackage{theorem}
\usepackage{latexsym,amssymb,amsfonts,amsmath}
\usepackage{graphicx}
\makeatletter
\@addtoreset{equation}{section}

\makeatother

\newtheorem{thm}{Theorem}[section]
\newtheorem{lem}[thm]{Lemma}
\newtheorem{cor}[thm]{Corollary}
\newtheorem{pro}[thm]{Proposition}

\theoremheaderfont{\scshape}

\newcommand{\ZM}{\mathbb{Z}}

\newcommand{\CM}{\mathbb{C}}

\newcommand{\ket}[1]{|#1\rangle}

\title{{\Large {\bf Periodicity for the Hadamard walk on cycles}}}
\author{
{\small Norio Konno$^{(1)}$, $\quad$ Yuki Shimizu$^{(2)}$, $\quad$ Masato Takei$^{(3)}$}\\
{\scriptsize Department of Applied Mathematics, 
Faculty of Engineering, 
Yokohama National University}\\
{\scriptsize 79-5 Hodogaya, Yokohama 240-8501, Japan}\\
{\scriptsize (1) e-mail: konno@ynu.ac.jp}\\
{\scriptsize (2) e-mail: shimizu-yoki-bx@ynu.jp}\\
{\scriptsize (3) e-mail: takei@ynu.ac.jp}\\
}

\date{\empty }
\begin{document}
\maketitle

\par\noindent
\begin{small}
\par\noindent
{\bf Abstract}. The present paper treats the period $T_N$ of the Hadamard walk on a cycle $C_N$ with $N$ vertices. Dukes (2014) considered the periodicity of more general quantum walks on $C_N$ and showed $T_2 =2, T_4=8, T_8=24$ for the Hadamard walk case. We prove that the Hadamard walk does not have any period except for his case, i.e., $N=2, 4, 8$. Our method is based on a path counting and cyclotomic polynomials which is different from his approach based on the property of eigenvalues for unitary matrix that determines the evolution of the walk.

\footnote[0]{
{\it Abbr. title:} Periodicity for the Hadamard walk on cycles
}
\footnote[0]{
{\it AMS 2000 subject classifications: }
60F05, 60G50, 82B41, 81Q99
}
\footnote[0]{
{\it PACS: } 
03.67.Lx, 05.40.Fb, 02.50.Cw
}
\footnote[0]{
{\it Keywords: } 
Quantum walk, Hadamard walk, periodicity, cycle
}
\end{small}

\setcounter{equation}{0}
\section{Introduction \label{intro}}
The quantum walk (QW) is the quantum counterpart of the classical random walk. QWs have been widely investigated for the last decade, mainly in connection with quantum information science. The reviews and books on QWs are, for example, Kempe \cite{Kempe2003}, Kendon \cite{Kendon2007}, Venegas-Andraca \cite{VAndraca2008, Venegas2013}, Konno \cite{Konno2008b}, Cantero et al. \cite{CanteroEtAl2013}, Manouchehri and Wang \cite{MW2013}, Portugal \cite{P2013}. The properties of QWs on graphs including cycles were studied by Aharonov et al. \cite{AharonovEtAl2001}. In this paper, we consider two-state QWs on a cycle $C_N$ with $N$ vertices, where $C_N = \{0,1, \ldots, N-1 \}$. In particular, we focus on the periodicity of the Hadamard walk on $C_N$.

From now on we present a brief definition of the general two-state QWs on $C_N$, which includes the Hadamard walk as a special case. The discrete-time QW is a quantum version of the classical random walk with additional degree of freedom called chirality. The chirality takes values left and right, and it means the direction of the motion of the walker. At each time step, if the walker has the left chirality, it moves one step to the left, and if it has the right chirality, it moves one step to the right. Let us define
\begin{align*}
\ket{L} = 
\begin{bmatrix}
1 \\
0  
\end{bmatrix},
\qquad
\ket{R} = 
\begin{bmatrix}
0 \\
1  
\end{bmatrix},
\end{align*}
where $L$ and $R$ refer to the left and right chirality states, respectively.  

The time evolution of the walk is determined by $U \in \mbox{\boldmath{U}}(2)$, where $\mbox{\boldmath{U}}(n)$ be the set of $n \times n$ unitary matrices and  
\begin{align*}
U =
\begin{bmatrix}
a & b \\
c & d
\end{bmatrix}.
\end{align*}
To define the dynamics of our model, we divide $U$ into two matrices:
\begin{eqnarray*}
P =
\begin{bmatrix}
a & b \\
0 & 0 
\end{bmatrix}, 
\quad
Q =
\begin{bmatrix}
0 & 0 \\
c & d 
\end{bmatrix},
\end{eqnarray*}
with $U =P+Q$. The important point is that $P$ (resp. $Q$) represents that the walker moves to the left (resp. right) at any position at each time step.

The QW considered here is 
\begin{align}
U = H = 
\frac{1}{\sqrt{2}}
\begin{bmatrix}
1 & 1 \\
1 & - 1 
\end{bmatrix} 
\label{akina}.
\end{align}
This model is called {\it the Hadamard walk} which has been extensively and deeply investigated in the study of QWs.

Let $\Psi_n$ denote the amplitude at time $n$ of the QW on $C_N$:  
\begin{align*}
\Psi_{n}
&= {}^T\! \left[\Psi_{n}^{L}(0),\Psi_{n}^{R}(0),\Psi_{n}^{L}(1),\Psi_{n}^{R}(1), \ldots, \Psi_{n}^{L}(N-1),\Psi_{n}^{R}(N-1) \right],
\\
&= {}^T\!\left[\Psi_{n}(0),\Psi_{n}(1), \cdots, \Psi_{n}(N-1) \right],
\\
&= {}^T\!\left[
\begin{bmatrix}
\Psi_{n}^{L}(0)\\
\Psi_{n}^{R}(0)\end{bmatrix},\begin{bmatrix}
\Psi_{n}^{L}(1)\\
\Psi_{n}^{R}(1)\end{bmatrix},
\ldots,
\begin{bmatrix}
\Psi_{n}^{L}(N-1)\\
\Psi_{n}^{R}(N-1)\end{bmatrix}
\right] \in(\mathbb{C}^{2})^{N},
\end{align*}
where $\CM$ denote the set of complex numbers, $T$ means the transposed operation, and $\Psi_n(x)={}^T \> [\Psi_n^{L}(x),\> \Psi_n^{R}(x)] \>\> (x \in C_N)$ is the amplitude at time $n$ and position $x$.

Now we introduce the following $2N \times 2N$ unitary matrix:
\begin{align*}\
U_N ^{(s)}=\begin{bmatrix}
O&P&O&O&\cdots&O&Q\\
Q&O&P&O&\cdots&O&O\\
O&Q&O&P&\cdots&O&O\\
O&O&Q&O&\cdots&O&O\\
\vdots&\vdots&\vdots&\vdots&\ddots&\vdots&\vdots\\
O&O&O&O&\cdots&O&P\\
P&O&O&O&\cdots&Q&O
\end{bmatrix}\;\;\;
\mbox{with} \;\;\;
O=\begin{bmatrix}
0&0\\
0&0
\end{bmatrix}.
\end{align*}
For $N=2$, following Dukes \cite{Dukes2014}, we put
\begin{align*}
U_2 ^{(s)} = 
\begin{bmatrix}
O & U \\
U & O 
\end{bmatrix}.
\end{align*}

Then the state of the QW at time $n$ is given by
\begin{align}
\Psi_{n}=(U_N^{(s)})^{n}\Psi_{0},
\label{sankeien}
\end{align} 
for any $n\geq0$.

Put $\mathbb{R}_{+}=[0,\infty)$. Here we introduce a map 
$\phi:(\mathbb{C}^{2})^{N} \rightarrow \mathbb{R}_{+}^{N}$
such that if
\begin{align*}
\Psi= {}^T\!\left[
\begin{bmatrix}
\Psi^{L}(0)\\
\Psi^{R}(0)\end{bmatrix},
\begin{bmatrix}
\Psi^{L}(1)\\
\Psi^{R}(1)\end{bmatrix},
\cdots,
\begin{bmatrix}
\Psi^{L}(N-1)\\
\Psi^{R}(N-1)\end{bmatrix}
\right] \in (\mathbb{C}^{2})^{N},
\end{align*}
then 
\begin{align*}
\phi(\Psi) = {}^T\! 
\left[
|\Psi^{L}(0)|^2 + |\Psi^{R}(0)|^2, 
|\Psi^{L}(1)|^2 + |\Psi^{R}(1)|^2,
\ldots, 
|\Psi^{L}(N-1)|^2 + |\Psi^{R}(N-1)|^2
\right] 
\in \mathbb{R}_{+}^{N}.
\end{align*}
That is, for any $x \in C_N$, 
\begin{align*}
\phi(\Psi) (x) = |\Psi^{L}(x)|^2 + |\Psi^{R}(x)|^2.
\end{align*}
Sometimes we identify $\phi(\Psi(x))$ with $\phi(\Psi) (x)$. Moreover we define the measure of the QW at position $x$ by
\begin{align*}
\mu(x)=\phi(\Psi(x)) \quad (x \in C_N).
\end{align*}

The probability that quantum walker at time $n$, $X_n= X_n ^{\varphi}$, starting from $0$ exists at position $x \in \ZM$ is defined by
\begin{align*}
P \left( X_n = x \right) =  P \left( X_n ^{\varphi} = x \right) =  \phi \left( \left( U^{(s)} \right)^n \Psi_0 ^{\varphi} \right) (x).
\end{align*}
Here the initial state $\Psi_0 ^{\varphi}$ is given by
\begin{align*}
\Psi_0 ^{\varphi}
= {}^T\!\left[
\begin{bmatrix}
\Psi_0 ^{L}(0)\\
\Psi_0^{R}(0)\end{bmatrix},
\begin{bmatrix}
\Psi_0^{L}(1)\\
\Psi_0^{R}(1)\end{bmatrix},
\cdots,
\begin{bmatrix}
\Psi_0^{L}(N-1)\\
\Psi_0^{R}(N-1)\end{bmatrix}
\right]
=
{}^T\!\left[
\varphi,
\begin{bmatrix}
0\\
0
\end{bmatrix},
\cdots,
\begin{bmatrix}
0\\
0
\end{bmatrix}
\right],
\end{align*}
where $\varphi = {}^T\![\alpha, \beta] \in \CM^2$ with $|\alpha|^2+|\beta|^2=1$.

We put
\begin{align*}
{\cal N} = \left\{ n \ge 1 : \left( U_N^{(s)} \right)^{n} = I_{2N} \right\}.
\end{align*}
If ${\cal N} \not= \emptyset$, the period $T_N (< \infty)$ is defined by $T_N = \min \cal{N}$. If ${\cal N} = \emptyset$, then we say that the QW does not have any period and write $T_N = \infty$. 

Let eigenvalues of $U_N^{(s)}$ be $\{\lambda_k : k=0,1, \ldots , 2N-1 \}$. Remark that $\displaystyle{\left(U_N^{(s)}\right)^{n} = I_{2N}}$ if and only if $\lambda_k^n =1 \> (k=0,1, \ldots , 2N-1).$

Dukes \cite{Dukes2014} studied periodicity of a class of two-state QWs on $C_N$ by using the property of eigenvalues $\lambda_k (k=0,1,\ldots, 2N-1)$ of $U_N ^{(s)}$: if period $T_N$ is finite, then $\lambda_j^{T_N}=1$ for any $j.$ As for the Hadamard walk case, he showed $T_2 =2, \> T_3 > 30, \> T_4=8, \> T_8=24.$ So we prove that $T_N = \infty$ except for $N=2,4,8$ (Theorem \ref{kahun3}).

The rest of this paper is organized as follows. In Sect. \ref{results}, we present results on our model. Sections \ref{kahun} and \ref{kahunda} are devoted to proofs of Lemma \ref{kahun1} and Theorem \ref{kahun3}, respectively. In Sect. \ref{sum}, we summarize our result and give a future problem.

\section{Results \label{results}}
This section gives our results. We begin with $N=3$ case. Then 
\begin{align*}
U_3 ^{(s)}=
\begin{bmatrix}
O&P&Q\\
Q&O&P\\
P&Q&O
\end{bmatrix}.
\end{align*}
So we have
\begin{align*}
\left(U_3 ^{(s)}\right)^2
&=
\begin{bmatrix}
PQ+QP&Q^2&P^2\\
P^2&PQ+QP&Q^2\\
Q^2&P^2&PQ+QP
\end{bmatrix},
\\
\left(U_3 ^{(s)}\right)^3
&=
\begin{bmatrix}
P^3+Q^3&PQP+QP^2+P^2Q&PQ^2+QPQ+Q^2P\\
PQ^2+QPQ+Q^2P&P^3+Q^3&PQP+QP^2+P^2Q\\
PQP+QP^2+P^2Q&PQ^2+QPQ+Q^2P&P^3+Q^3 
\end{bmatrix}.
\end{align*}
Let $A(k,l) \> (1 \le k, l \le m)$ denote the $(k,l)$ component of matrix $A$. For example, $\displaystyle{\left( U_3 ^{(s)} \right)^3} (1,2)=PQP+QP^2+P^2Q$.

In order to compute $\displaystyle{\left( U_N ^{(s)} \right)^n (k,l)}$, we use nice relations: $P^2 = aP, \> Q^2=dQ.$ Moreover we introduce the following $2 \times 2$ matrices, $R$ and $S$:
\begin{align*}
R=
\begin{bmatrix}
c & d \\
0 & 0 
\end{bmatrix}, 
\quad
S=
\begin{bmatrix}
0 & 0 \\
a & b 
\end{bmatrix}.
\end{align*}
Then we obtain the next table of products of matrices, $P, \> Q, \> R,$ and $S$:
\par
\
\par
\begin{center}
\begin{tabular}{c|cccc}
  & $P$ & $Q$ & $R$ & $S$  \\ \hline
$P$ & $aP$ & $bR$ & $aR$ & $bP$  \\
$Q$ & $cS$ & $dQ$& $cQ$ & $dS$ \\
$R$ & $cP$ & $dR$& $cR$ & $dP$ \\
$S$ & $aS$ & $bQ$ & $aQ$ & $bS$ 
\label{pqrs}
\end{tabular}
\end{center}
\par
\
\par\noindent
where $PQ=bR$, for example. In particular, for the Hadamard walk case, we have 
\par
\
\par
\begin{center}
\begin{tabular}{c|rrrr}
  & \multicolumn{1}{c}{$P$} & \multicolumn{1}{c}{$Q$} & \multicolumn{1}{c}{$R$} & \multicolumn{1}{c}{$S$}  \\ \hline
$P$ & $P/\sqrt{2}$ & $R/\sqrt{2}$ & $R/\sqrt{2}$ & $P/\sqrt{2}$  \\
$Q$ & $S/\sqrt{2}$ & $-Q/\sqrt{2}$& $Q/\sqrt{2}$ & $-S/\sqrt{2}$ \\
$R$ & $P/\sqrt{2}$ & $-R/\sqrt{2}$& $R/\sqrt{2}$ & $-P/\sqrt{2}$ \\
$S$ & $S/\sqrt{2}$ & $Q/\sqrt{2}$ & $Q/\sqrt{2}$ & $S/\sqrt{2}$ \\ 
\end{tabular}
\end{center}
\par
\
\par\noindent
This path counting method was introduced and intensively studied by \cite{Konno2002, Konno2005}. Using this relation, we compute
\begin{align*}
\left( U_3 ^{(s)} \right)^3 (1,2)=PQP+QP^2+P^2Q=bcP+abR+acS= \left( \frac{1}{\sqrt{2}} \right)^2 \left( P + R + S \right). 
\end{align*}
Similarly we have
\begin{align*}
\left( U_3 ^{(s)} \right)^4 (1,2)
&=Q^3P + P^4 + PQ^3 +QPQ^2+ Q^2PQ \\
&=a^3P+bd^2R+(cd^2 + bcd + bcd)S \\
&= \left( \frac{1}{\sqrt{2}} \right)^3 \left\{ P + R + (1-1-1) S \right\}. 
\end{align*}
Moreover we write the number of paths corresponding to $\displaystyle{\left( U_N ^{(s)} \right)^n (k,l)}$ by $w (N, n ; (k,l))$, e.g., $w (3, 2 ; (1,2))=1, \> w (3, 3 ; (1,2))=3, \> w (3, 4 ; (1,2))=5$. In general, the property of paths easily implies
\begin{lem}
For any $N \ge 2, \> n \ge 1$, $w (N, n ; (k,k))$ is even for $k \in \{1,2, \ldots, N\}$ and $w (N, n ; (1,l))=w (N, n ; (1,N-(l-2)))$ for $l \in \{2,3, \ldots, [(N/2)+1] \}$, where $[x]$ is the integer part of real number $x$. 
\label{salad1}
\end{lem}

We should note that $P, \> Q, \> R,$ and $S$ form an orthogonal basis of the vector space of $2 \times 2$ matrices with respect to the trace inner product $\langle A | B \rangle = $ tr$(A^{\ast}B)$. Thus if there exist $c_p, c_q, c_r, c_s \in \CM$ such that 
\begin{align*}
c_p P + c_q Q + c_r R + c_s S = O_2,
\end{align*}
then $c_p=c_q=c_r=c_s=0$. Thus we see that for any $N \ge 2, \> n \ge 1$ and $k \in \{1,2, \ldots, N\}$, if $w (N, n ; (k,l))$ is odd, then $\displaystyle{\left(U_N ^{(s)}\right)^n} (k,l) \not= O_2.$ Therefore combining this property with Lemma \ref{salad1}, we obtain the following lemma which is one of the key results of our method.
\begin{lem}
If there exists $n \ge 1$ such that $\displaystyle{\left(U_N ^{(s)}\right)^n = I_{2N}}$, then $w (N, n ; (k,l))$ is even for any $k,l \in \{1,2, \ldots, N\}$. 
\label{salad2}
\end{lem}


To count the number of paths, we introduce the adjacency matrix $A_N$ of $C_N$:
\begin{align*}\
A_N = 
\begin{bmatrix}
0&1&0&\cdots&0&1\\
1&0&1&\cdots&0&0\\
0&1&0&\cdots&0&0\\
\vdots&\vdots&\vdots&\ddots&\vdots&\vdots\\
0&0&0&\cdots&0&1\\
1&0&0&\cdots&1&0
\end{bmatrix}.
\end{align*}
For example, in $N=3$ case, we have
\begin{align*}
A_3=
\begin{bmatrix}
0&1&1\\
1&0&1\\
1&1&0
\end{bmatrix},
\quad
(A_3)^2=
\begin{bmatrix}
2&1&1\\
1&2&1\\
1&1&2
\end{bmatrix},
\quad
(A_3)^3=
\begin{bmatrix}
2&3&3\\
3&2&3\\
3&3&2
\end{bmatrix}.
\end{align*}

Moreover we introduce another $N \times N$ matrix $B_N ^{(n)}$ whose component $B_N ^{(n)} (k,l)$ is equal to $(A_N)^n (k,l) \> ({\rm mod} \> 2)$. For $N=3$ case, we get
\begin{align*}
B_3 ^{(1)}=B_3 ^{(2)} = B_3 ^{(3)} = \cdots = 
\begin{bmatrix}
0&1&1\\
1&0&1\\
1&1&0
\end{bmatrix}.
\end{align*}
By using notation $B_N^{(n)} (k,l)$, Lemma \ref{salad2} can be rewritten as
\begin{lem}
If there exists $n \ge 1$ such that $\displaystyle{\left(U_N ^{(s)}\right)^n = I_{2N}}$, then $B_N^{(n)} (k,l)=0$ for any $k,l \in \{1,2, \ldots, N\}$. 
\label{salad3}
\end{lem}

On the other hand, we have the following result. 
\begin{lem}
For any odd number $N (\ge 3)$, we have $B_N^{(n)} (k,l)=1$ for some distinct $k,l \in \{1,2, \ldots, N\}$. 
\label{kahun1}
\end{lem}
The proof will appear in Sect. \ref{kahun}.  Combining Lemma \ref{salad3} with Lemma \ref{kahun1} immediately gives 
\begin{pro}
For any odd number $N (\ge 3)$, we have $T_N = \infty$.
\label{kahun2}
\end{pro}

By using Proposition \ref{kahun2} and the property of cyclotomic polynomials, we obtain the following main result.
\begin{thm}
For any $N$ except for $N=2,4,8$, we have $T_N = \infty$.
\label{kahun3}
\end{thm}
The proof will appear in Sect. \ref{kahunda}. We should remark that Higuchi et al. \cite{HiguchiEtAl} investigated the periodicity of the Szegedy walk on graphs, e.g., the complete graphs, by using a method based on the property of cyclotomic polynomials. On the other hand, we consider the periodicity of the Hadamard walk on cycles by using not only cyclotomic polynomials but also the path counting for the walk.

Combining Dukes' result, $T_2 =2, T_4=8, T_8=24$, with our Theorem \ref{kahun3} gives immediately 
\begin{thm}
For any $N \ge 2$,
\begin{eqnarray*}
T_N = \left\{ 
\begin{array}{cl}
2,
&
(N=2),
\\
8,
& 
(N=4),
\\
24,
& 
(N=8),
\\
\infty,
& 
(N \not= 2,4,8).
\end{array} \right.
\end{eqnarray*}
\label{kahun4}
\end{thm}


We should note that for the classical random walk in which the walker moves one step to the left with probability $p$ and to the right with probability $q$ with $p+q=1 \> (p,q \in [0,1])$, the eigenvalues $\{ \lambda_k : k=0,1, \ldots, N-1 \}$ of the corresponding transition matrix are given by
\begin{align*}
\lambda_k = \cos \left( \frac{2 k \pi}{N} \right) + i \> (q-p) \sin \left( \frac{2 k \pi}{N} \right) \quad (k=0,1, \ldots, N-1).
\end{align*}

Therefore we see that for any $N \ge 2$,
\begin{eqnarray*}
T_N = \left\{ 
\begin{array}{cl}
N,
&
(p=0, 1),
\\
\infty,
& 
(p \in (0,1)),
\end{array} \right.
\end{eqnarray*}
since $\lambda_k =e^{2 \pi i k/N} \>(p=0), \> e^{-2 \pi i k/N} \>(p=1),$ and $|\lambda_1|<1 \> (0<p<1).$

From now on we briefly review previous results on the Hadamard walk on $C_N$. To do so, we define the time-averaged measure $\overline{\mu}_{n}$ at time $n$ and the limit measure $\overline{\mu}_{\infty}$ for the Hadamard walk on $C_N$ by
\begin{align*}
\overline{\mu}_{n} (x) 
&= \frac{1}{n} \sum_{k=0}^{n-1} P(X_k =x), 
\\
\overline{\mu}_{\infty} (x) &= \lim_{n \to \infty} \frac{1}{n} \sum_{k=0}^{n-1} P(X_k =x)
\end{align*}
for any $x \in C_N$. Aharonov et al. \cite{AharonovEtAl2001} proved that the time-averaged limit measure $\overline{\mu}_{\infty}$ is uniform for odd $N$, that is, $\overline{\mu}_{\infty} (x) = 1/N \> (x \in C_N)$, independent of the initial state.

Bednarska et al. \cite{BednarskaEtAl2003} considered the Hadamard walk on $C_N$ with even $N$. They obtained the eigenvalues and eigenvectors of $U_N^{(s)}$ and gave an explicit formula of $\overline{\mu}_{\infty}$ starting from a single vertex for any $N$. By using the formula, they showed that $\overline{\mu}_{\infty}$ is uniform for $N=2,4$. Moreover they found that $\overline{\mu}_{\infty}$ is very sensitive to the arithmetric properties of $N$.

Bednarska et al. \cite{BednarskaEtAl2004} reported examples for three different kinds of behaviour of the total variation distance between a uniform measure and the time-averaged measure $\overline{\mu}_{n}$ for the Hadamard walk on $C_N$ with even $N$.

From Theorem \ref{kahun4}, we have
\begin{cor}
For $N=2,4,8$, 
\begin{align*}
\overline{\mu}_{\infty} = \frac{1}{T_N} \sum_{n=0}^{T_N-1} \mu_n.
\end{align*}
\end{cor}

\section{Proof of Lemma \ref{kahun1} \label{kahun}}
Before we move to the proof, we consider $N=5$ case. In this case, $B_5^{(1)}(=A_5)$ is given by
\begin{align*}
B_5^{(1)} = 
\begin{bmatrix}
0&1&0&0&1\\
1&0&1&0&0\\
0&1&0&1&0\\
0&0&1&0&1\\
1&0&0&1&0
\end{bmatrix}.
\end{align*}
Then we would like to show that for any $n \ge 1$, there exists $(k,l)$ with $k \not= l$ such that $B_5^{(n)} (k,l)=1$ by induction. 

When $n=1$, we see immediately $B_5^{(1)} (1,2)=1$. 

Next we consider $n=m$ case. We put 
\begin{align}
\left[ B_5^{(m)} (1,1), B_5^{(m)} (1,2), B_5^{(m)} (1,3), B_5^{(m)} (1,4), B_5^{(m)} (1,5) \right]
=\left[0, c_1, c_2, c_2, c_1 \right].
\label{okazu1}
\end{align}
Here we should remark that Lemma \ref{salad1} implies $B_5^{(m)} (1,1)=0$ and $c_1 = B_5^{(m)} (1,2)=B_5^{(m)} (1,5), \> c_2 = B_5^{(m)} (1,3)=B_5^{(m)} (1,4)$ for any $m \ge 1$. We assume that when $n=m$, the statement holds. That is, $(c_1, c_2) \not= (0,0).$ We consider $n=m+1$. We assume that 
\begin{align}
\left[ B_5^{(m+1)} (1,1), B_5^{(m+1)} (1,2), B_5^{(m+1)} (1,3), B_5^{(m+1)} (1,4), B_5^{(m+1)} (1,5) \right] =\left[0, 0, 0, 0, 0 \right].
\label{okamoto}
\end{align}
Then Eq. \eqref{okazu1} implies
\begin{align}
\left[ B_5^{(m+1)} (1,1), B_5^{(m+1)} (1,2), B_5^{(m+1)} (1,3), B_5^{(m+1)} (1,4), B_5^{(m+1)} (1,5) \right]
=\left[0, c_2, c_1+c_2, c_1 + c_2, c_2 \right].
\label{taro}
\end{align}
Combining Eq. \eqref{okamoto} with Eq. \eqref{taro} gives 
\begin{align*}
c_2=c_1+c_2=0.
\end{align*}
Thus we have $c_1=c_2=0$. This contradicts the assumption for $n=m$, i.e., $(c_1, c_2) \not= (0,0).$ Therefore we see that there exists $(1,l)$ with $l \in \{2,3,4,5\}$ such that $B_5^{(m+1)} (1,l)=1$. So Lemma \ref{kahun1} is valid for $N=5$.

We can extend this argument to general odd number $N=2M+1$ as follows.

When $n=1$, we easily see $B_N^{(1)} (1,2)=1$. Next we consider $n=m$. In a similar way, we put 
\begin{align}
\left[ B_N^{(m)} (1,1), B_N^{(m)} (1,2), \ldots , B_N^{(m)} (1,N) \right]
=\left[0, c_1, c_2, \ldots , c_{M-1}, c_M, c_M, c_{M-1}, \ldots, c_2, c_1 \right].
\label{okazu2}
\end{align}
We assume that when $n=m$, the statement holds. That is, $(c_1, c_2, \ldots , c_M) \not= (0, 0, \ldots , 0)$. We consider $n=m+1$. We assume that 
\begin{align}
\left[ B_N^{(m+1)} (1,1), B_N^{(m+1)} (1,2), \ldots, B_N^{(m+1)} (1,N) \right] =\left[0, 0, \ldots, 0 \right].
\label{gokamoto}
\end{align}
Then Eq. \eqref{okazu2} gives
\begin{align}
&
\left[ B_N^{(m+1)} (1,1), B_N^{(m+1)} (1,2), \ldots, B_N^{(m+1)} (1,N) \right] 
\nonumber
\\
& \qquad =\left[0, c_2, c_1+c_3, c_2 + c_4, \ldots, c_{M-2}+c_M, c_{M-1}+c_M, \right.
\nonumber
\\
& \qquad \qquad \qquad \left. c_{M-1}+c_M, c_{M-2}+c_M, \ldots, c_2 + c_4,c_1+c_3, c_2 \right].
\label{gtaro}
\end{align}
From Eq. \eqref{gokamoto} and Eq. \eqref{gtaro}, we obtain  
\begin{align*}
c_2=c_1+c_3= c_2 + c_4= \cdots = c_{M-2}+c_M=c_{M-1}+c_M=0.
\end{align*}
Thus we have $c_1=c_2= \cdots = c_M=0$. This contradicts the assumption for $n=m$, i.e., $(c_1, c_2, \ldots , c_M) \not= (0, 0, \ldots , 0)$. Therefore we see that there exists $(1,l)$ with $l \in \{2, \ldots ,N \}$ such that $B_N^{(m+1)} (1,l)=1$, and the proof is completed.

\section{Proof of Theorem \ref{kahun3} \label{kahunda}}
First we introduce cyclotomic polynomials: $F_1 (\lambda) = \lambda -1$, and for $n \ge 2$, 
\begin{align*}
F_n (\lambda) = \prod_{\scriptstyle 1 \le k \le n-1: \atop \scriptstyle {\rm gcd}(k,n)=1} \left( \lambda - \exp \left(\frac{2 \pi i k}{n} \right) \right),
\end{align*}
where ${\rm gcd} (n_1, n_2, \ldots , n_k)$ denotes the greatest common divisor of $(n_1, n_2, \ldots , n_k)$. 

Before we move to a proof of Theorem \ref{kahun3}, we give another proof of Dukes' result, $T_2 =2, \> T_4=8, \> T_8=24$, by using cyclotomic polynomials. By definition of $U_N^{(s)}$, we have
\begin{align}
\det \left( \lambda I_{2N} - U_N^{(s)} \right) = \prod_{k=0}^{N-1} \left\{ \lambda^2 + i \sqrt{2} \sin \left(\frac{2 \pi i k}{N} \right) \lambda - 1 \right\},
\label{smidori}
\end{align}
see \cite{BednarskaEtAl2003,BednarskaEtAl2004}, for example. From Eq. \eqref{smidori}, we compute 
\begin{align*}
\det \left( \lambda I_{4} - U_2^{(s)} \right) 
&= F_1(\lambda)^2 \> F_2(\lambda)^2,
\\
\det \left( \lambda I_{8} - U_4^{(s)} \right) 
&= F_1(\lambda)^2 \> F_2(\lambda)^2 \> F_8(\lambda),
\\
\det \left( \lambda I_{16} - U_8 ^{(s)} \right) 
&= F_1(\lambda)^2 \> F_2(\lambda)^2 \> F_8(\lambda) \> F_{12}(\lambda)^2.
\end{align*}
Then we have the desired conclusion:
\begin{align*}
T_2 = {\rm lcm} (1,2) =2, \quad T_4 = {\rm lcm} (1,2,8) =8, \quad T_8 = {\rm lcm} (1,2,8,12) =24, 
\end{align*}
where ${\rm lcm} (n_1, n_2, \ldots , n_k)$ denotes the least common multiple of  $(n_1, n_2, \ldots , n_k)$.

Next we give another proof of $N=3$ case of Proposition \ref{kahun2} by using cyclotomic polynomials. That is, we prove $T_3 = \infty$. From Eq. \eqref{smidori}, we calculate 
\begin{align*}
\det \left( \lambda I_{6} - U_3^{(s)} \right) = F_1(\lambda) \> F_2(\lambda) \> G(\lambda),
\end{align*}
where 
\begin{align*}
G(\lambda) =  \lambda^4- \frac{\lambda^2}{2}+1.
\end{align*}
On the other hand, it is known that there are only four cyclotomic polynomials with degree 4 as follows: 
\begin{align*}
F_5(\lambda) & =\lambda^4+\lambda^3+\lambda^2+\lambda+1, \quad F_8(\lambda)=\lambda^4+1, \\
F_{10}(\lambda) & =\lambda^4-\lambda^3+\lambda^2-\lambda+1, \quad F_{12}(\lambda)=\lambda^4-\lambda^2+1.
\end{align*}
Thus, we confirm that $G(\lambda)$ is not a cyclotomic polynomial and conclude that $T_{3}= \infty.$

From now on, we move to a proof of Theorem \ref{kahun3}. First we consider odd $N$ case. Then we have
\begin{pro}
For any odd $N$, there exist $m (N), r_1, r_2, \ldots, r_{m(N)} \ge 1$ such that 
\begin{align*}
\det \left( \lambda I_{2N} - U_N^{(s)} \right) = \prod_{j=1}^{m(N)} F_{r_j} (\lambda) \times G(\lambda),
\end{align*}
where $G(\lambda)$ is not a cyclotomic polynomial.
\label{midorida}
\end{pro}
The proof is that if we do not have such a $G(\lambda)$, then $T_N = {\rm lcm} (r_1, r_2, \ldots , r_{m(N)})< \infty$ and this contradicts Proposition \ref{kahun2}. 

Moreover, we consider $N=2^n \times M$ case, where $n \ge 1$ and $M$ is an odd number. By Eq. \eqref{smidori}, we see that there exists a polynomial $H(\lambda)$ such that
\begin{align*}
&\det \left( \lambda I_{2N} - U_N^{(s)} \right) \\
&= \prod_{k=0}^{2^n \times M -1} \left\{ \lambda^2 + i \sqrt{2} \sin \left(\frac{2 \pi i k}{2^n \times M} \right) \lambda - 1 \right\} \\
&= \left\{ \lambda^2 + i \sqrt{2} \sin \left(\frac{2 \pi i \times 2^n \times 0}{2^n \times M} \right) \lambda - 1 \right\} \times  \left\{ \lambda^2 + i \sqrt{2} \sin \left(\frac{2 \pi i \times 2^n \times 1}{2^n \times M} \right) \lambda - 1 \right\} 
\\ 
& \times \cdots \times \left\{ \lambda^2 + i \sqrt{2} \sin \left(\frac{2 \pi i \times 2^n \times (M-1)}{2^n \times M} \right) \lambda - 1 \right\} \times H(\lambda) \\
&= \det \left( \lambda I_{2M} - U_M^{(s)} \right) \times H(\lambda). 
\end{align*}
From Proposition \ref{midorida}, we see that there exist $m (M), r_1, r_2, \ldots, r_{m(M)} \ge 1$ such that 
\begin{align*}
\det \left( \lambda I_{2N} - U_N^{(s)} \right) = \prod_{j=1}^{m(M)} F_{r_j} (\lambda) \times G_M (\lambda) \times H(\lambda),
\end{align*}
where $G_M (\lambda)$ is not a cyclotomic polynomial. So we have $T_N = \infty$ for $N=2^n \times M$, where $n \ge 1$ and $M$ is an odd number.

Therefore it is enough to deal with $N=2^n \> (n \ge 4)$ cases, since $T_2=2, \> T_{2^2}=8, \> T_{2^3} = 24$. 

For $N=2^4 =16$ case, we obtain
\begin{align}
\det \left( \lambda I_{2^5} - U_{2^4} ^{(s)} \right) 
&= F_1(\lambda)^2 \> F_2(\lambda)^2 \> F_8(\lambda) \> F_{12}(\lambda)^2 
\nonumber \\
& \times \left( \lambda^4- \frac{\lambda^2}{2}+1 \right)^2 \> \left( \lambda^4- \frac{3 \lambda^2}{2}+1 \right)^2.
\label{midori1}
\end{align}
Thus, as in the case of $N=3$, we obtain $T_{2^4}= \infty$.

For $N=2^n \> (n \ge 5)$, we see that there exists a polynomial $G_N (\lambda)$ such that 
\begin{align}
\det \left( \lambda I_{2^{n+1}} - U_{2^n}^{(s)} \right) = \det \left( \lambda I_{2^5} - U_{2^4} ^{(s)} \right) \times G_N (\lambda). 
\label{midori2}
\end{align}
Combining Eq. \eqref{midori1} with Eq. \eqref{midori2} implies that there exist  $m (N), r_1, r_2, \ldots, r_{m(N)} \ge 1$ and a polynomial $G^{\ast}_N (\lambda)$ such that 
\begin{align*}
\det \left( \lambda I_{2^{n+1}} - U_{2^n}^{(s)} \right) = \prod_{j=1}^{m(N)} F_{r_j} (\lambda) \times G^{\ast}_N (\lambda).
\end{align*}
We should note that $G^{\ast}_N (\lambda)$ contains a factor $\lambda^4- \lambda^2/2+1$. So $G^{\ast}_N (\lambda)$ is not a cyclotomic polynomial. Therefore we conclude that $T_N = \infty$ for any $N=2^n \> (n \ge 5)$.

\section{Summary \label{sum}}
In this paper, we proved that the period $T_N = \infty$ except with $N=2, 4, 8$ for the Hadamard walk on $C_N$. On the other hand, $T_2=2, \> T_4=8, \> T_8 = 24$ was previously shown by Dukes \cite{Dukes2014} in 2014. Our method is based on a path counting and cyclotomic polynomials which is different from his approach based on the property of eigenvalues for $U_N^{(s)}$. An implementation of a Hadamard-like QW on $C_N$ was proposed by Moqadam et al. \cite{MoqadamEtAl2014} by using optomechanical systems. We hope that our result is helpful in building new quantum algorithms. Chou and Ho \cite{ChouHo2014} investigated numerically the asymptotic behaviour of space-inhomogeneous QWs on $\ZM$, where $\ZM$ is the set of integers. Their model is defined by a periodic quantum coin $U_x (x \in \ZM)$ given by $H$ or $I_2$, where $I_2$ is the $2 \times 2$ identity matrix, e.g., $U_x =H$ for $x=0$ (mod $N$), $U_x =I_2$ for $x \not=0$ (mod $N$) with $N \ge 2$. They discussed localization of the QWs, so one of the interesting future problems is to consider the periodicity of space-inhomogeneous QWs on $C_N$.

\par
\
\par\noindent
{\bf Acknowledgments.} We would like to thank Hyun Jae Yoo, Chul Ki Ko, Takeshi Kajiwara, Seiya Yoshida, Yuto Minowa, Kei Saito for useful discussions. This work was partially supported by the Grant-in-Aid for Scientific Research (C) of Japan Society for the Promotion of Science (Grant No.24540116).

\par
\
\par

\begin{small}
\bibliographystyle{jplain}

\end{small}

\end{document}